\documentclass[usenatbib]{mnras}
\usepackage{newtxtext,newtxmath}
\usepackage[T1]{fontenc}
\usepackage{ae,aecompl}

\usepackage{graphicx}	
\usepackage{amssymb}
\usepackage{amsmath}	
\usepackage{amssymb}	

\title{Two planetary systems with transiting Earth-size and super-Earth planets orbiting late-type dwarf stars}

\author[E.D. Alonso et al.]{E. D\'iez Alonso$^{1}$,
J. I. Gonz\'alez Hern\'andez$^{2,3}$,
S. L. Su\'arez G\'omez$^{4}$,
D. S. Aguado$^{2}$,
\newauthor C. Gonz\'alez Guti\'errez$^{1}$,
A. Su\'arez Mascare\~no$^{5}$,
A. Cabrera-Lavers$^{2,6}$,
\newauthor J. Gonz\'alez-Nuevo$^{4}$,
B. Toledo--Padr\'on$^{2,3}$,
J. Gracia$^{7}$,
F. J. de Cos Juez$^{1}$\thanks{E-mail: fjcos@uniovi.es}, R. Rebolo$^{2,3,8}$
\\
$^{1}$Department of Exploitation and Exploration of Mines, University of Oviedo, Oviedo, Spain\\
$^{2}$Instituto de Astrof\'isica de Canarias, E--38205 La Laguna, Tenerife, Spain\\
$^{3}$Universidad de La Laguna, Dpto. Astrof\'isica, E--38206 La Laguna, Tenerife, Spain\\
$^{4}$Departamento de F\'isica, Universidad de Oviedo, C. Federico Garc\'ia Lorca 18, E-33007, Oviedo, Spain\\
$^{5}$Observatoire Astronomique de l\textquotesingle Universit\'e de G\`eneve, 1290 Versoix, Switzerland\\
$^{6}$GRANTECAN, Cuesta de San Jos\'e s/n, E-38712, Bre\~na Baja, La Palma, Spain\\
$^{7}$Department of Construction and Manufacturing Engineering, University of Oviedo, Oviedo, Spain\\
$^{8}$Consejo Superior de Investigaciones Cient\'ificas, Spain\\
}

\date{Accepted 2018 June 04. Received 2018 April 25}

\pubyear{2018}

\begin{document}
\label{firstpage}
\pagerange{\pageref{firstpage}--\pageref{lastpage}}
\maketitle

\begin{abstract}

We present two new planetary systems found around cool dwarf stars with data from the K2 mission. The first system was found in K2-239 (EPIC 248545986), characterized in this work as M3.0V and observed in the $14^{th}$ campaign of K2. It consists of three Earth-size transiting planets with radii of 1.1, 1.0 and 1.1 $R_{\oplus}$, showing a compact configuration with orbital periods of 5.24, 7.78 and 10.1 days, close to 2:3:4 resonance. The second was found in K2-240 (EPIC 249801827), characterized in this work as M0.5V and observed in the $15^{th}$ campaign. It consists of two transiting super-Earths with radii 2.0 and 1.8 $R_{\oplus}$ and orbital periods of 6.03 and 20.5 days. The  equilibrium temperatures of the atmospheres of these planets are estimated to be in the range of 380-600 K and the amplitudes of signals in transmission spectroscopy are estimated at $\sim$ 10 ppm.

\end{abstract}




\section{INTRODUCTION}
Low-mass stars are primary targets in the search for Earth-size planets and in the study of their properties.
Low-mass stars (0.1 $M_{\odot}$ < M < 0.6 $M_{\odot}$) account for 70\% of the stellar population in the Milky Way \citep{1994AJ....108.1437H}, meaning they have a hugely significant impact in the overall statistics of planets in the Galaxy.
Exoplanets with close-in orbits tend to be terrestrial when the mass of the star decreases \citep{2012ApJS..201...15H}, with an average of $\sim$ 0.5 Earth-size rocky planet with $P_{\rm orb}$ < 50 days around each low-mass star \citep{2015ApJ...807...45D}.

Transiting Earth-size planets induce deeper dimmings in the light-curve of low mass stars and stronger radial velocity signals than in more massive stars. Temperate planets orbit closer and have shorter orbital periods, so it is easier to detect planets in the habitable zone (orbital range in which a planet's atmosphere can warm the surface to allow surface liquid water) \citep{2016Natur.536..437A,2017Natur.542..456G}. Signals in transit transmission spectroscopy \citep{2002ApJ...568..377C} are also stronger for stars with a small radius, so planets orbiting near bright low-mass stars are also suitable for atmospheric characterization \citep{2014Natur.505...69K}.

Detecting transiting planetary systems is of great value in terms of estimating the mass and density of their planets measuring transit timing variations \citep{2017Natur.542..456G}, which are stronger for compact systems in resonances. These systems are also suitable for testing the formation scenarios from the study of resonances that could be the result of migrations \citep{2005MNRAS.363..153P}.

Until now, the Kepler mission \citep{2010Sci...327..977B} has been the most successful facility detecting exoplanets by the transit method. Since the beginning of 2014, Kepler has been on its second mission (K2) \citep{2014PASP..126..398H}, monitoring different fields near the ecliptic plane for $\sim$80 days. K2 has found many exoplanet candidates \citep{2015ApJ...800...59V,2016ApJS..226....7C,2018MNRAS.476L..50D,2018AJ....155..124H} in each observation campaign.

Campaign 14 was conducted between May 31st and August $19^{th}$ 2017, centering on the Leo and Sextant area (central coordinates $\alpha$=10:42:44, $\delta$=+06:51:06). Campaign 15 ran between August 23rd and November $20^{th}$ 2017, observing the area towards the constellation of Scorpius (central coordinates $\alpha$=15:34:28, $\delta$=-20:04:44).

In this study we present the detection of two planetary system during these campaings. The first consists of three Earth-size transiting planets orbiting K2-239 (EPIC 248545986) ($\alpha$=10:42:22.633, $\delta$=+04:26:28.86), observed in long cadence mode during campaign 14. The second consists of two transiting super-Earths orbiting K2-240 (EPIC 249801827 ($\alpha$=15:11:23.907, $\delta$=-17:52:30.78), observed in long cadence mode during campaign 15.

\section{Spectroscopic and Photometric Data}
\subsection{Stellar characterization: K2-239}
\begin{figure}
	\includegraphics [width=0.45\textwidth]{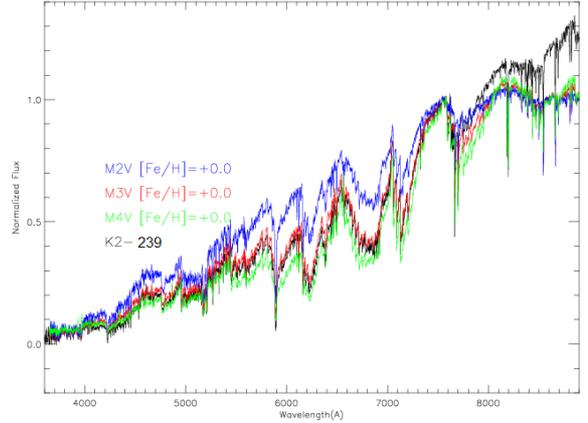}
    \caption{OSIRIS spectrum of K2-239 compared with reference spectra of M2.0V - M4.0V stars. Best fit is obtained for M3V star with $\mathrm{[Fe/H]} \sim 0$. All spectra are normalized at $\lambda$ = 7575 {\AA}.}
    \label{fig:Especosiris}
\end{figure}

On March $13^{th}$ 2018 we obtained spectra of K2-239 with the OSIRIS camera-spectrograph \citep{2000SPIE.4008..623C} of the 10.4 m Gran Telescopio Canarias (GTC), located at Observatorio Roque de los Muchachos in La Palma (Canary Islands, Spain). Three medium-resolution spectra ($\lambda/\delta \lambda\sim2500$) in each of the BVRI bands were reduced in the standard manner, flux calibrated, telluric corrected, and finally combined into a single spectrum (see Fig.~\ref{fig:Especosiris}).

The spectrum was compared with SDSS/BOSS reference spectra of M-type stars from \cite{2017ApJS..230...16K}. The comparison was made with the HAMMER code \citep{2007AJ....134.2398C}, obtaining the best fit for a M3V  star with $\mathrm{[Fe/H]} \sim 0$. The relative intensity of the NaI lines at 5890 and 8180 {\AA} rule out the possibility of the star being giant, while the relative depth of the strong molecular bands of TiO at 7000-7300 {\AA} points to $\mathrm{[Fe/H]} \sim 0$.
\cite{2015A&A...577A.132M}, working with measurements of spectral index from 
HARPS spectra, conclude $T_{\rm eff}$ $\sim$ 3450 $\pm$ 50 K for M3V stars, which is in agreement with our estimates of the stellar parameters.
Figure ~\ref{fig:Especosiris}  plots our comparison of the OSIRIS spectrum of K2-239 with reference spectra from M2.0V to M4.0V stars.

We computed the stellar parameters from J,H,V,K magnitudes listed in Table~\ref{tab:stellar_parameters}, applying the empirical relationships established by \cite{2013AJ....145...52M,2015ApJ...804...64M} and \cite{2013ApJS..208....9P}, using the tabulated stellar parameters from \cite{2013ApJS..208....9P} and the Mass - Luminosity relation for Main-sequence M dwarfs from \cite{2016AJ....152..141B}. All the parameters are listed in Table~\ref{tab:stellar_parameters}.

Taking $m_{V} = 14.549 \pm 0.040$ (Table~\ref{tab:stellar_parameters}) and $M_{V} = 11.09 \pm 0.10$ from \cite{2013ApJS..208....9P} tabulated parameters, we estimate a distance to K2-239 of $49 \pm 3$ pc.

We measured a radial velocity from the OSIRIS
spectrum $v_{r}=-8.5 \pm 1.5$ kms$^{-1}$, which combined with the estimated distance and the proper motions $\mu_{\alpha} = -41.0 \pm 3.9$ [mas/yr] and $\mu_{\delta} =10.5\pm 8.4$ [mas/yr], results in the velocity components listed in Table~\ref{tab:stellar_parameters}. From the probability distributions of \cite{2006MNRAS.367.1329R}, we derive that K2-239 is a member of the Galactic thin disk.

\subsection{Stellar characterization: K2-240}
K2-240 has been observed by the Radial Velocity Experiment (RAVE) \citep{2006AJ....132.1645S}. RAVE's DR5 \citep{2017AJ....153...75K} presents data from medium-resolution spectra (R $\sim$ 7500) covering the Ca-triplet region (8410-8795 \AA). From RAVE's DR5 we find for K2-240 (RAVE J151123.9-175231) $T_{\rm eff} = 3800 \pm 87$ K and $\log g = 4.50 \pm 0.17$, confirming that K2-240 is a cool dwarf star.

We repeated exactly the same analysis followed for K2-239 to derive the stellar parameters accurately, obtaining the parameters listed in Table~\ref{tab:stellar_parameters}. These parameters are consistent with K2-240 being a M0.5V star.

We also note that a very clear rotation signal is present in the light curve from K2. From a Lomb-Scargle \citep{1982ApJ...263..835S} analysis we estimate $P_{\rm rot} = 10.8 \pm 0.1$ d. 

From RAVE's radial velocity $v_{r}=0.20 \pm 1.56$  kms$^{-1}$, our estimated distance of $d = 70 \pm 3$ pc, and proper motions $\mu_{\alpha} = -53.6 \pm 1.5$ [mas/yr] and $\mu_{\delta} =-49.8\pm 1.0$ [mas/yr], we compute velocity components listed in Table~\ref{tab:stellar_parameters}. From the probability distributions of \cite{2006MNRAS.367.1329R}, we derive that K2-240 is a member of the Galactic thin disk.

\begin{table}
	  
	\centering
	\caption{Stellar parameters for K2-239 and K2-240}
	\label{tab:stellar_parameters}
    \resizebox{9 cm}{!} {
	\begin{tabular}{lccc}
		       
        Parameter&K2-239&K2-240&Source\\
        \hline
        \hline
        V [mag]& $ 14.549\pm 0.040 $&$13.392\pm0.010$&(1)\\
        R [mag]& $ 13.906\pm 0.020 $&$12.804\pm0.010$&(1)\\
        I [mag]& $ 12.718\pm 0.030 $&$11.994\pm0.050$&(1)\\
        J [mag]& $ 10.781\pm 0.026 $&$10.394\pm0.027$&(2)\\
        H [mag]& $ 10.192\pm 0.021 $&$9.745\pm0.024$&(2)\\
        K [mag]& $  9.971\pm 0.021 $&$9.560\pm0.023$&(2)\\
        $T_{\rm eff}$ [K]& $ 3420\pm 18$&$ 3810\pm 17$&(3)\\
        $\mathrm{[Fe/H]}$& $ -0.1\pm 0.1$&$-0.1\pm 0.1$&(3)\\              
       	Radius [$R_{\odot}$] &$0.36 \pm 0.01$&$0.54 \pm 0.01$&(3)\\
		Mass [$M_{\odot}$]& $0.40\pm0.01$&$0.58\pm0.01$&(3)\\
        Luminosity [$L_{\odot}$]& $0.016 \pm 0.001$&$0.053 \pm 0.002$&(3)\\
        $\log g$ [cgs]& $4.9\pm0.1$&$4.7\pm0.1$&(3)\\
        $P_{\rm rot}$ [d]&--&$10.8 \pm 0.1$&(3)\\
        Distance [pc] &$49 \pm 3$ &$70 \pm 3$&(3)\\
        $\mu_{\alpha}$ [mas/yr]&$-41.0 \pm 3.9$&$-53.6 \pm 1.5$&(1)\\
        $\mu_{\delta}$ [mas/yr]&$10.5\pm 8.4$&$-49.8\pm 1.0$&(1)\\
        U, V, W  [km/s]&-6.8, 4.2, -10.2&-5.4, -23.6, -1.7&(3)\\
       &&\\
       \multicolumn{3}{l}{(1) UCAC4 \citep{2013AJ....145...44Z}.}\\
         \multicolumn{3}{l}{(2) 2MASS \citep{2003tmc..book.....C}.}\\
         \multicolumn{3}{l}{(3) This work.}\\
            
	\end{tabular}
}
\end{table}

\subsection{K2 photometric data}

\begin{figure*}
	\includegraphics [width=0.9\textwidth]{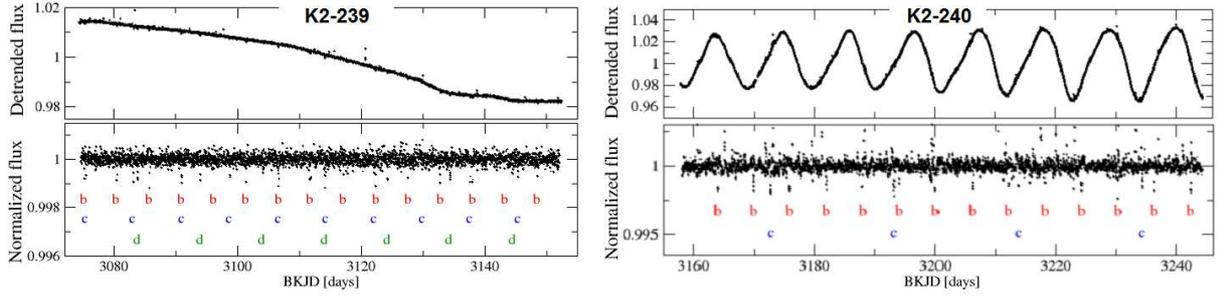}
    \caption{K2 detrended (top) and normalized (bottom) light curves for K2-239 (left) and K2-240 (right). Characters b, c and d show times of observed transits for planets in each system.}
    \label{fig:lightcurves}
\end{figure*}

\begin{figure}
	\includegraphics [width=0.40\textwidth]{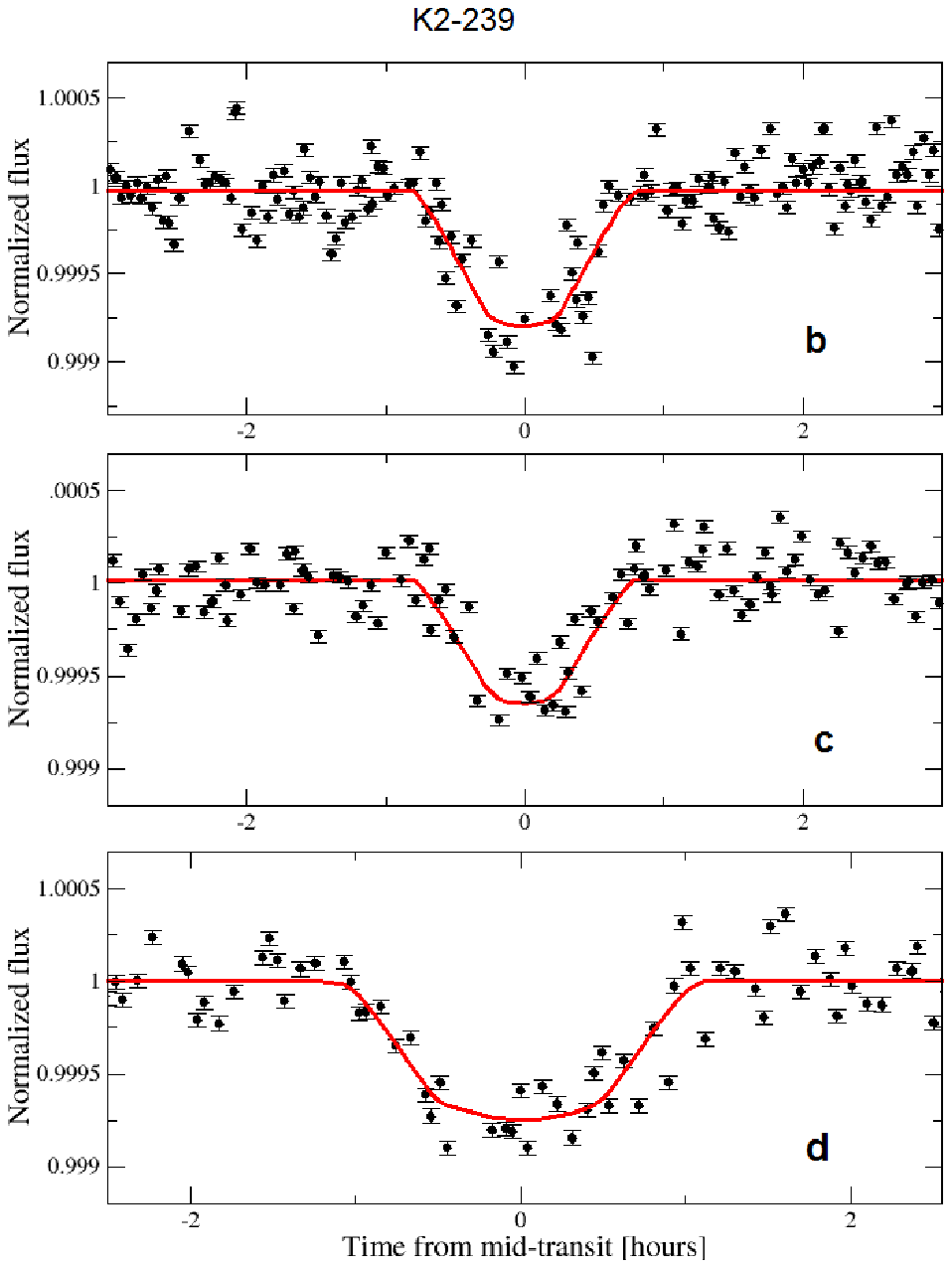}
    \caption{Phase-folded light curves corresponding to planets b (top), c (middle), and d (bottom) in the K2-239 system. Solid curves represent best model fits obtained by MCMC.}
    \label{fig:transits248545986}
\end{figure}
\begin{figure}
	\includegraphics [width=0.40\textwidth]{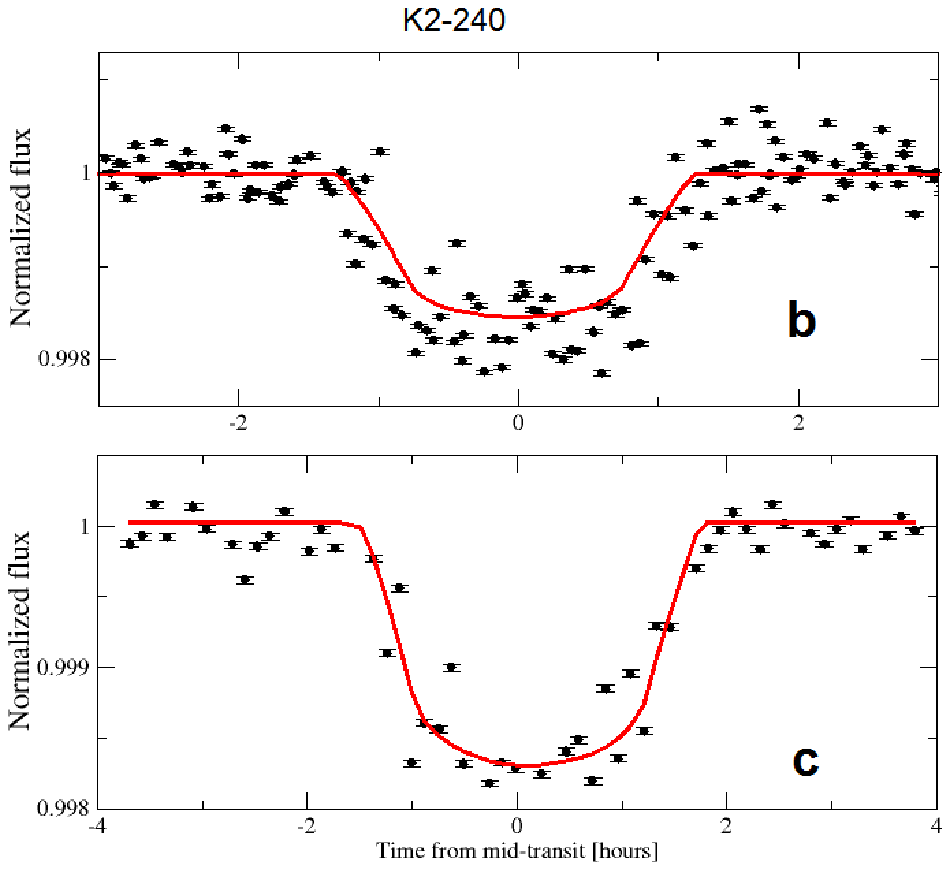}
    \caption{Phase-folded light curves corresponding to planets b (top) and c (bottom) in the K2-240 system. Solid curves represent best model fits obtained by MCMC.}
    \label{fig:transits249801827}
\end{figure}

We followed the work of \cite{2014PASP..126..948V} to analyze the K2 corrected photometry of our target stars, detrending stellar variability with a spline fit and searching for periodic signals using the Box Least Squares (BLS) method \citep{2002A&A...391..369K} on attained data. This analysis shows three transit signals with periods 5.240$\pm$0.001 (b), 7.775$\pm$0.001 (c) and 10.115$\pm$0.001 (d) days in K2-239 and two transit signals with periods 6.034$\pm$0.001 (b) and 20.523$\pm$0.001 (c) days in K2-240 (Fig.~\ref{fig:lightcurves}).

We performed MCMC analysis on each phase-folded transit (Figs.~\ref{fig:transits248545986} and~\ref{fig:transits249801827}) to estimate the planetary parameters, fitting models from \cite{2002ApJ...580L.171M} with the Exofast package \citep{2013PASP..125...83E}. For each data point, the light curve was resampled 10 times uniformly spaced over the 29.5-minute long cadence of K2 and averaged, following  \cite{2010MNRAS.408.1758K}. For the calculations we set the values of $T_{\rm eff}$, $\log g$, and $\mathrm{[Fe/H]}$ listed in Table~\ref{tab:stellar_parameters}, and orbital periods listed above. We also worked with the assumption of eccentricity $e = 0$, valid for transiting planets in a multi-planetary system \citep{2015ApJ...808..126V}.

The planets in the K2-239 system have estimated radii $1.1\pm 0.1$ $R_{\oplus}$ (b), $1.0\pm 0.1$ $R_{\oplus}$ (c) and $1.1\pm 0.1$ $R_{\oplus}$ (d), orbital periods of 5.242$\pm$0.001 days (b), 7.775$\pm$0.001 days (c) and 10.115$\pm$0.001 days (d), and semimajor axis $0.0441\pm 0.0008$ AU (b), $0.0576\pm 0.0009$ AU (c) and $0.0685\pm 0.0012$ AU (d).

The planets in the K2-240 system have estimated radii $2.0^{+0.2}_{-0.1}$ $R_{\oplus}$ (b) and $1.8^{+0.3}_{-0.1}$ $R_{\oplus}$ (c), orbital periods of 6.034$\pm$0.001 days (b) and 20.523$\pm$0.001 days (c), and semimajor axis $0.0513\pm 0.0009$ AU (b), $0.1159\pm 0.0020$ AU (c).
Table~\ref{tab:planetparams} summarizes all the parameters obtained for the planets.

\subsection{False positives analysis}

\begin{figure}
	
	\includegraphics [width=0.40\textwidth]{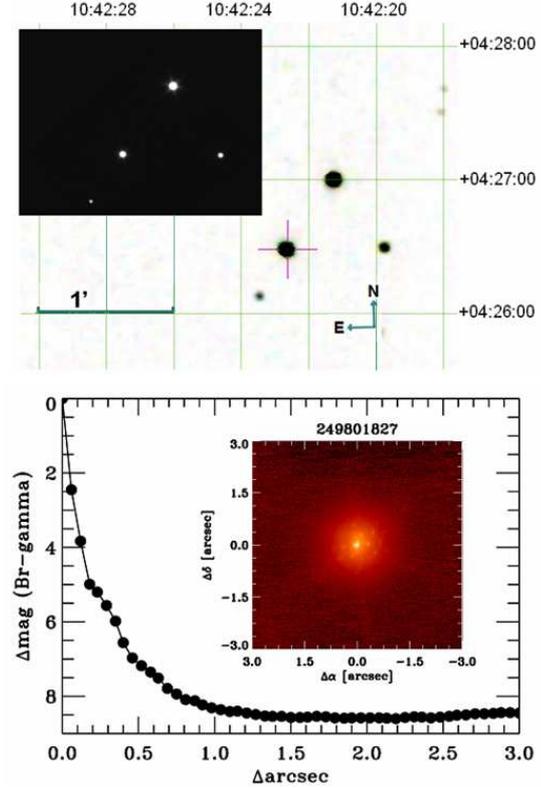}
    \caption{Top panel: OSIRIS/GTC image taken of K2-239 field with seeing 0.6 arc seconds in the i-band Sloan filter, superimposed on the 2MASS image of the field. Bottom panel: contrast curve and AO image of K2-240 acquired with the NIRC2 instrument at the 10 m Keck-2 telescope.  }
    \label{fig:falsepositives}
\end{figure}

We acquired images of K2-239 with the OSIRIS camera-spectrograph on March $13^{th}$ 2018. Night conditions were rather good, and data were collected under photometric conditions, a dark moon, and with an average seeing of 0.7 arc seconds. For broadband imaging, a series of 10 x 1 sec in Sloan i filter was obtained. Bias correction, flat fielding and bad pixel masking were done using standard procedures, and the images were finally aligned (see Fig.~\ref{fig:falsepositives}, top panel). Analyses of final image exclude companions at 0.6 arc seconds with $\delta$mag < 5.0 and at 3 arc second with $\delta$mag < 10.

In the same way, images from POSS-I \citep{1963bad..book..481M} (year 1953) and 2MASS \citep{2003tmc..book.....C} (year 1998, see Fig.~\ref{fig:falsepositives}, top panel) do not show background sources at the current star position.

At ExoFOP--K2 \footnote{https://exofop.ipac.caltech.edu/k2/} an AO image of K2-240 is available, acquired with the NIRC2 instrument at the 10 m Keck 2 telescope (Maunakea, Hawaii). The image excludes companions at 0.2 arc seconds with $\delta$mag < 5.0 and at 1 arc second with $\delta$mag < 8.3 (Fig~\ref{fig:falsepositives}, bottom panel).

Non-detection of blended objects in these images and the extremely low probability of multiple false positives as shown by \cite{2011ApJS..197....8L} confirm the planetary origin of  transit signals in K2-239 and K2-240.

\begin{table*}
	\centering
   
	\caption{Parameters for planets in the K2-239 and K2-240 systems.}
	\label{tab:planetparams}
	
    \resizebox{16cm}{3.8cm} {
    \begin{tabular}{lccccccc}
    	
        Planet Parameters&K2-239 b&K2-239 c&K2-239 d& K2-240 b&K2-240 c\\
        \hline
        \hline
        Orbital period (P) [d]& $5.240\pm0.001$&$7.775\pm0.001$&$10.115\pm0.001$&$6.034\pm0.001$&$20.523\pm0.001$\\
         \hline
        Semi-major axis (a) [AU]& $0.0441\pm 0.0008$&$0.0576\pm 0.0009$&$0.0685\pm0.0012$&$0.0513\pm0.0009$&$0.1159\pm0.0020$\\
         \hline
        
Radius ($R_{p}$) [$R_{\oplus}$] & $1.1\pm0.1$&$1.0\pm0.1$&$1.1\pm0.1$&$2.0^{+0.2}_{-0.1}$&$1.8^{+0.3}_{-0.1}$\\
		 \hline
        Mass ($M_{p}$) [$M_{\oplus}$]  & $1.4\pm0.4$&$0.9\pm0.3$&$1.3\pm0.4$&$5.0^{+0.5}_{-0.2}$&$4.6^{+0.7}_{-0.3}$\\
        \hline
        
        Equilibrium Temperature ($T_{eq}$) [K] & $502^{+22}_{-18}$&$427^{+24}_{-19}$&$399^{+18}_{-15}$&$586^{+24}_{-18}$&$389^{+19}_{-17}$\\
        &&&&&\\
   		Transit Parameters&K2-239 b&K2-239 c&K2-239 d& K2-240 b&K2-240 c\\
        \hline
        \hline
       
        Epoch (BKJD) [days]&3075.191&3083.860&3075.381&3163.825&3172.722\\
        \hline
 
        Radius of planet in stellar radii ($R_{p}$/$R_{*}$)&$0.0259^{+0.0013}_{-0.0012}$ &$0.0241^{+0.0016}_{-0.0014}$&$0.0255\pm0.0012$&$0.0362 \pm 0.0014$&$0.0313 \pm 0.0024$\\
        \hline

        Semi major axis in stellar radii         (a/$R_{*}$)&$24.6^{+1.9}_{-2.0}$ &$34.0^{+3.3}_{-3.5}$&$38.8^{+2.9}_{-3.3}$&$21.1^{+1.3}_{-1.6}$&$48.0 \pm 4.4$\\
        \hline
    
        Linear limb-darkening coeff ($u_{1}$)&$0.344^{+0.058}_{-0.055}$ &$0.344^{+0.049}_{-0.050}$&$0.345\pm0.056$&$0.402^{+0.068}_{-0.067}$&$0.411^{+0.071}_{-0.067}$\\
        \hline
        
        Quadratic limb-darkening coeff ($u_{2}$)&$0.373\pm0.053$ &$0.374^{+0.051}_{-0.050}$&$0.371^{+0.051}_{- 0.052}$&$0.308^{+0.062}_{- 0.063}$&$0.312^{+0.061}_{- 0.064}$\\
        \hline
               
        Inclination (i) [deg]&$88.99^{+0.68}_{-0.87}$ &$88.77^{+0.70}_{-0.57}$&$89.43^{+0.38}_{-0.45}$&$89.26^{+0.51}_{-0.64}$&$89.66^{+0.22}_{-0.26}$\\
        \hline 
       
        Impact Parameter (b)&$0.44^{+0.33}_{-0.29}$ &$0.73^{+0.24}_{-0.39}$&$0.39\pm0.25$&$0.28^{+0.20}_{-0.19}$&$0.29^{+0.18}_{-0.19}$\\     
        \hline
        
        Transit depth ($\delta$)&$0.00067^{+0.00007}_{-0.00006}$ &$0.00058^{+0.00008}_{-0.00006}$&$0.00065\pm0.00006$&$0.00131\pm 0.00010$&$0.00098^{+0.00016}_{-0.00015}$\\
        \hline
                       
        Total duration ($T_{14}$) [d]&$0.0611^{+0.0067}_{-0.013}$&$0.052^{+0.013}_{-0.027}$&$0.0766^{+0.0058}_{-0.0073}$&$0.0897^{+0.0046}_{-0.0038}$&$0.133^{+0.011}_{-0.010}$\\

	\end{tabular}

}

\end{table*}

\section{Discussion and conclusions}

Assuming the planet radii listed in Table~\ref{tab:planetparams}, and the mean density for planets satisfying $R_{p} \leq 1.5 R_{\oplus}$ from \cite{2014ApJ...783L...6W}, we obtain $M_{b}=1.4\pm 0.4$ $M_{\oplus}$, $M_{c}=0.9\pm 0.3$ $M_{\oplus}$, $M_{d}=1.3\pm 0.4$ $M_{\oplus}$ for planets b, c, and d, respectively in the K2-239 system. Adopting $M_{p} \ll M_{*}$, circular orbits and $\sin i\sim$1, we computed induced semi-amplitudes in stellar velocity variations of 0.9 ms$^{-1}$ for planet b, 0.5 ms$^{-1}$ for planet c and 0.7 ms$^{-1}$ for planet d, well-suited  for radial velocity monitoring with ultra-stable spectrographs such as ESPRESSO \citep{2014AN....335....8P,2017arXiv171105250G} at the VLT.

The amplitude of the signal in transit transmission spectroscopy can be estimated as $\frac{R_{p} \cdot {h_{\rm eff}}}{(R_{*})^{2}}$ \citep{2016Natur.533..221G} with $h_{\rm eff}$ the effective atmospheric height. $h_{\rm eff}$ is related to the atmospheric scale height H = K$\cdot$T/$\mu$$\cdot$g (K Boltzmann's constant, T atmospheric temperature, $\mu$ atmospheric mean molecular mass, g surface gravity). Assuming $h_{\rm eff}$ = 7$\cdot$H \citep{2010ApJ...716L..74M} for a transparent volatile dominated atmosphere ($\mu$ = 20) with 0.3 Bond albedo, we found amplitudes in transit transmission spectroscopy of 1.2$\cdot 10^{-5}$ (b), 1.1$\cdot 10^{-5}$ (c) and $10^{-5}$ (d).

We used the Mercury package \citep{1999MNRAS.304..793C} to simulate and test the evolution and stability of the system for $10^{6}$ years. We simulated using Bulirsch -- Stoer integrator, adopting circular orbits and masses from the mass-radius relation. We do not find significant changes in the eccentricity or in the inclination of the orbits, showing a dynamically stable system.

To estimate the masses for the planets of the K2-240 system we used the mass-radius relation from \cite{2014ApJ...783L...6W} for planets satisfying 
$1.5\leq R_{p}/R_{\oplus}\leq4$, obtaining $M_{b}=5.0^{+0.5}_{-0.2}$ $M_{\oplus}$, $M_{c}=4.6^{+0.7}_{-0.3}$ $M_{\oplus}$. 
Under the assumption of $M_{p} \ll M_{*}$, circular orbits and $\sin i\sim$1, we computed induced semi-amplitudes in stellar velocity variations of 2.5 ms$^{-1}$ for planet b and 1.5 ms$^{-1}$ for planet c.
With the same assumptions as in the previous section, we estimate amplitudes in transit transmission spectroscopy of 1.2$\cdot 10^{-5}$ (b) and 6.6$\cdot 10^{-6}$ (c).

We also tested the stability of K2-240 system with the Mercury package as described in the previous section. Again our simulations point towards a dynamically stable system.

The planetary systems presented in this work, with equilibrium temperatures estimated in the range 380-600 K, are suitable targets for incoming facilities; Plato, monitoring in shorter cadence mode, could reveal transit timing variations that allow accurate planetary masses to be estimated. The James Webb Telescope could find signs of planetary atmospheres. Ultra-stable spectrographs such as ESPRESSO at VLT, could also carry out radial velocity follow-up,  so these are promising targets to improve our understanding of compact Earth-sized planetary systems (K2-239) and super-Earth systems on the rocky-gaseous boundary (EPIC K2-240). 

\section*{Acknowledgements}
EDA, CGG and JCJ acknowledge Spanish ministry project AYA2017-89121-P. JIGH, BTP, DSA and RRL acknowledge the Spanish ministry project MINECO AYA2014- 56359-P. JIGH also acknowledges financial support from the Spanish Ministry of Economy and Competitiveness (MINECO) under the 2013 Ram\'on y Cajal program MINECO RYC-2013-14875. ASM acknowledges financial support from the Swiss National Science Foundation (SNSF).
JGN and SLSG acknowledge financial support from the I+D 2015 project
AYA2015- 65887-P (MINECO/FEDER). JGN also acknowledges financial support
from the Spanish MINECO for a Ram\'on y Cajal fellowship (RYC-2013-13256).

Based on observations made with the Gran Telescopio Canarias (GTC), installed in the Spanish Observatorio del Roque de los Muchachos of the Instituto de Astrof\'isica de Canarias, in the island of La Palma.






\bsp	
\label{lastpage}
\end{document}